# Customer Data Sharing Platform: A Blockchain-Based Shopping Cart

Ajay Kumar Shrestha
*Department of Computer Science*
*University of Saskatchewan*
Saskatoon, Canada
ajay.shrestha@usask.ca

Sandhya Joshi
*ComIT*
Saskatoon, Canada
sandhyajoshi01@gmail.com

Julita Vassileva
*Department of Computer Science*
*University of Saskatchewan*
Saskatoon, Canada
jiv@cs.usask.ca

*Abstract*— We propose a new free-ecommerce platform with blockchains that allows customers to connect to the seller directly, share personal data without losing control and ownership of it and apply it to the domain of shopping cart. Our new platform provides a solution to four important problems: private payment, ensuring privacy and user control, and incentives for sharing. It allows the trade to be open, transparent with immutable transactions that can be used for settling any disputes. The paper presents a case study of applying the framework for a shopping cart as one of the enterprise nodes of MultiChain which provides trading in ethers controlled by smart contracts and also collects users' profile data and allows them to receive rewards for sharing their data with other business enterprises. It tracks who shared what, with whom, when, by what means and for what purposes in a verifiable fashion. The user data from the repository is converted into an open data format and shared via stream in the blockchain so that other nodes can efficiently process and use the data. The smart contract verifies and executes the agreed terms of use of the data and transfers digital tokens as a reward to the customer. The smart contract imposes double deposit collateral to ensure that all participants act honestly.

## I. INTRODUCTION

Over the years, there has been a revolutionary advancement in the filed of technological innovation and related research on collaborative approaches for sharing users' data among enterprises [1]. User data is collected by different parties, for example, companies offering apps, e-commerce sites, online social networking sites and others, whose primary motive is to have enhanced business model while giving optimal services to their customers. However, the collection of user data is associated with serious privacy and security issues [2]. A flexible mechanism for obtaining and renewing consent for data use and sharing is required that provides appropriate and meaningful incentives for users to capitalize from data sharing and ensures transparency for users to be aware of which of their dataset has been accessed, by whom, for what purpose and under what conditions [3].

This paper, therefore, presents a novel platform to ensure privacy and user control, and incentives for sharing while addressing the issue of private payment in the context of online shopping cart system. Here is a list of objectives for conducting this research work.

- To create decentralized e-commerce experience for customers.
- To enable companies to increase trust in their products and supply chains.
- To offer direct payment with native Ethereum tokens thereby enabling privacy and confidentiality.
- To create a proof of the existence to every transaction.
- To give the users full transparency over who accesses their data, when and for what purpose,
- To enable companies to share customers' data among others in the consortium network.
- To provide incentives to customers in real-time for sharing their data.

This novel blockchains based platform has 3-tier shopping cart application employing *Spring Boot* and *React* as the main building technologies, that allows users to shop online using ether with all the transactions stored in the blockchain eliminating the trust, and to get incentivised upon permitting to share their own data as stated in the smart contracts.

## II. BACKGROUND

Blockchain technology holds promise to transform data management in many domains. Although blockchain was developed as a platform for virtual currency, the applications of the technology have since quickly evolved to numerous fascinating use cases [4]. In contrast to the centralized system, blockchain technology can be totally transparent to the users and very promising to incentivize users for data sharing [5]. It also naturally supports building up incentives for users to share their data, in terms of rewards (micro-payments or credits) encoded in the smart contracts.

We have used MultiChain as a private blockchain that provides the privacy and control required in an easy to configure and deploy package [6]. Ethereum, on the other hand, is an open-source, public blockchain to create decentralized applications (dapps) where users interact with the online services in a distributed peer-to-peer manner that takes place on a censorship-proof foundation [7]. Ethereum has "ether" (ETH) as its virtual currency which is used to pay a transaction fee and to provide a primary liquidity layer for exchanging digital assets. Ethereum also offers a technology called smart contracts, which are instances of contracts deployed on the blockchain although the term was originally coined earlier [8] in the context of electronic commerce protocols between strangers on the Internet. A smart contract stores the rules which negotiate the terms of the contract, automatically verifies the contract, and executes the agreed terms.

## III. RELATED WORKS AND CURRENT WORK

The paper [1] presented in an ACM conference in Big Data depicted a decentralized storage architecture with blockchain technology. Another paper [9] on sharing users' personal data with Blockchain demonstrated a distributed data sharing architecture with blockchain in the travel domain. The paper also presented an evaluation of the performance of the model by measuring the latency and memory consumption with three test scenarios that mostly affect the user experience.

Similarly, another paper [3] presented at the International Conference on Blockchain in Seattle in March 2018 is the first in the area of sharing research data. The paper presented a prototype for sharing research data. It allows researchers a proper way of creating the proof of existence and, tracking the sharing of their extensive research data and samples, and receiving incentives in real-time in the form of digital tokens (with monetary value) or attribution for their research work (credit, citation, or a collaboration offer) or both. Its implementation and evaluation were done in the paper [10] by measuring the transaction cost for smart contracts deployment. There are a few open-source decentralized marketplace projects such as *OpenBazaar* [11] that support peer to peer transactions with cryptocurrencies. However, it does not have any provision for offering incentives to users for sharing their data in the digital marketplace.

Therefore, this paper is different in the context that it has used an online marketplace domain and presented the work for a shopping cart using both the Ethereum Smart contracts and MultiChain blockchain to offer a novel platform for customers to make payment with digital tokens and receive incentives for sharing their personal data among enterprises. The platform is proposed to combine a payment mechanism through ethers and a mutual agreement between customers and sellers via smart contracts. It automatically registers the immutable timestamped metadata about transactions which can be useful to settle any possible disputes among stakeholders in the future. Furthermore, the enterprises share their customer data among their consortium network through a secure permissioned blockchain network that keeps track of who shared what, to whom, when, for what purpose and under what condition.

## IV. SYSTEM DEVELOPMENT

The system has a 3 tier restful maven shopping cart application that uses Spring Boot as the backend technology with certain dependencies such as Lombok, starter-data-jpa allowing opinionated configurations yet connecting to MYSQL database and React which is built with node.js runtime environment covering the frontend part written in the ES6 version of JavaScript providing the dynamic refreshment of pages upon changes with flexible, responsive and intuitive user experience. Furthermore, the system has the permissioned MultiChain as a solution to both on-chain and off-chain data storage, encryption, hashing and tracking of data, together with Ethereum for access control and enabling transactions with ethers.

Fig. 1 presents the interaction among the customer (data provider) and other e-commerce apps (data consumers) of the proposed blockchain-based user-controlled data sharing scheme. This platform allows customer to do purchase items using ethers. It also offers the usage of the MultiChain for the on-chain and off-chain storage of customer data at the shopping cart node. The delivery of the key is done using asymmetric cryptography, which is provided with the MultiChain commands. MultiChain nodes handle hashing and encrypting the customer data, storing the encrypted file off-chain, committing the hash of the file on the blockchain, searching the required data, verifying the data and delivering the data. All the participants in the system with their Ethereum account addresses are in the MultiChain network. The shopping cart node puts data in the local storage and the MultiChain enables encrypting it, creates a transaction with their hashes and commits it into the blockchain. The data consumer node subscribes to the stream searching for the items (data) and finds the off-chain item with the help of their metadata and hashes. It then places the hash portion in its retrieval queue that queries the data in the P2P network.

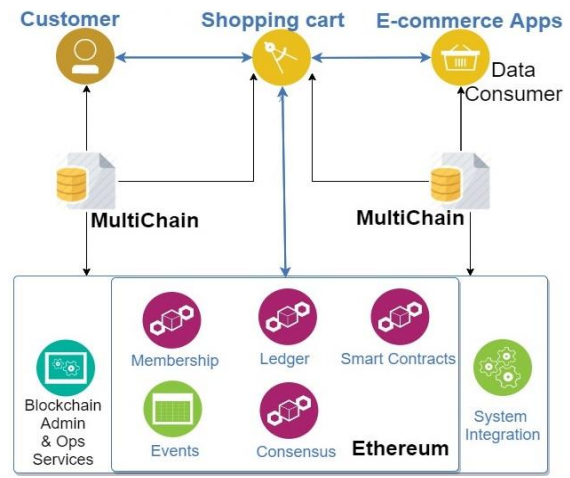

Fig. 1. Blockchains based shopping cart with a data sharing platform

The node which possesses the data signature (data owner) responds to the query. At this point, the smart contracts get triggered and with their successful execution, the tokens are transferred from the data consumer's Ethereum address to the customer's account while delivering the requested data (with verified hashes) to the local storage of the consumer node using the same path. Ethereum is for access control and enabling transactions with ethers.

The performance evaluation and the user study are important measures to determine the success of the developed system. Similar studies have already been done in our previous system with similar functionalities [9], [10], [2]. Since there are two blockchain platforms in our system, we considered measuring latency and memory consumption parameters for the private network because the data sharing is performed in that network, which requires very low latency for optimal performance and the storage delay might also play a role in the increased latency and poor performance.

## V. CONCLUSION

We presented a novel platform for the online shopping cart based on user-controlled privacy and data-sharing policies encoded in smart contracts. The use of blockchains enables the building up a verifiable record of the provenance, accountability of access and incentives for customers to share their data, in terms of rewards. Data sharing is done in the private blockchain network and there is no storage of data on the public blockchain, so data deletion rule in a compliant manner can delete each category of data. Moreover, there is transparency and control over which enterprise has access to the relevant personal data when and for what purpose.